\documentclass[12pt,epsfig]{article}
\usepackage[xdvi]{graphicx}

\newcommand{\beq}{\begin{equation}}
\newcommand{\eeq}{\end{equation}}
\newcommand{\bea}{\begin{eqnarray}}
\newcommand{\eea}{\end{eqnarray}}

\def\e2sig{e^{-2r\sigma}}

\begin{document}
\setlength{\baselineskip}{18pt}
\begin{titlepage} 
\begin{flushright}
KOBE-TH-08-02
\end{flushright}
\vspace*{10mm}
\begin{center}{\large\bf Finite Gluon Fusion Amplitude 
in the Gauge-Higgs Unification}
\end{center}
\vspace*{10mm}
\begin{center}
{\bf Nobuhito Maru}
\end{center}
\vspace*{0.2cm}
\begin{center}
${}^{a}${\it Department of Physics, Kobe University, 
Kobe 657-8501, Japan}
\end{center}
\vspace*{2cm}
\begin{abstract}
We show that the gluon fusion amplitude 
 in the gauge-Higgs unification scenario is finite 
 in any dimension regardless of its nonrenormalizability. 
This result is supported by the fact that the local operator describing 
 the gluon fusion process is forbidden by the higher dimensional gauge invariance. 
We explicitly calculate the gluon fusion amplitude 
 in an arbitrary dimensional gauge-Higgs unification model 
 and indeed obtain the finite result. 
\end{abstract}
\end{titlepage}


Gauge-Higgs unification \cite{Manton}-\cite{Hosotani3} is one of the leading candidates 
 solving the gauge hierarchy problem without supersymmetry. 
In this scenario, the Higgs field is regarded as extra spatial components 
 of the higher dimensional gauge field, which immediately forbids the local Higgs mass term 
 thanks to the higher dimensional gauge invariance. 
Then, the nonlocal finite Higgs mass is obtained by Wilson loop dynamics 
 regardless of the nonrenormalizability of the theory, which has been explicitly verified 
 in various models \cite{mass}-\cite{HMTY}. 
The finite Higgs mass is very predictive since it is independent of the cutoff scale of the theory. 
This opens up a possibility to solve the gauge hierarchy problem without relying on supersymmetry 
 and a huge number of interesting works on the gauge-Higgs unification have been reported 
 from various viewpoints \cite{LM}-\cite{PW}.

It is natural to ask whether there are any other finite physical observables 
 in the gauge-Higgs unification. 
If there are, we can naively guess that it is the physical observables composed of the gauge field 
 and the Higgs field, which are nontrivially transformed under the higher dimensional gauge symmetry. 
Along this line of thought, 
 the divergence structure of S and T parameters was investigated \cite{LM}. 
Unfortunately, they are divergent in theories more than five dimensions 
 as anticipated from the power counting argument, 
 but a particular linear combination of them was found to be finite in six dimension case \cite{LM}. 
Surprisingly enough, the anomalous magnetic moment of the fermion was recently shown 
 to be finite in any dimension although not only the gauge and Higgs fields 
 but also fermions are included in the local operator describing the anomalous magnetic moment \cite{ALM}.

The Large Hadron Collider (LHC) experiments are just about to start 
 and the various collider signatures predicted from the various extensions to the Standard Model 
 have recently been studied. 
The present author and N. Okada also investigated 
 the processes of the gluon fusion Higgs boson production 
 and the two photon decay of Higgs boson, which are main processes in case of the light Higgs boson,  
 in a five dimensional gauge-Higgs unification model \cite{MO} (see also \cite{Falkowski, LR}). 
We have found that the amplitudes of these processes become finite 
 and are independent of the cutoff scale of the theory. 
This result itself is very natural from the power counting argument 
 since the local operators relevant for the gluon fusion and the two photon decay 
 are dimension six operators and our calculations were done in a five dimensional model. 
However, we can guess from the structure of Kaluza-Klein (KK) mass spectrum and Yukawa coupling 
 that this finiteness nature holds true in any space-time dimensions. 
Also, the above mentioned local operator is composed of only the gauge and Higgs fields, 
 which suggests that the gluon fusion and two photon decay amplitudes are finite.

In this paper, we show that the gluon fusion amplitude in the gauge-Higgs unification scenario 
 is finite in any space-time dimension regardless of its nonrenormalizability. 
Here we take a toy model of $(D+1)$ dimensional $SU(3)$ gauge-Higgs unification 
 with a triplet fermion identified with third generation quarks. 
The extra spatial dimension is compactified on an orbifold $S^1/Z_2$. 
Although this model is a toy model in that the Weinberg angle is too large 
 $\sin^2 \theta_W = 3/4$, the top quark is massless and 
 the bottom quark mass is equal to a $W$ boson mass and so on, 
 it is enough to show the finiteness of the gluon fusion amplitude and 
 to avoid inessential complications. 

The action of our model is given by 
\bea
S = \int d^Dx dy \left[ 
-\frac{1}{2} \mbox{Tr}  (F_{MN}F^{MN}) 
+ i\bar{\Psi} D\!\!\!\!/_{D+1} \Psi
\right],
\label{lagrangian}
\eea
where the $D+1$ dimensional covariant derivative is defined as 
 $D\!\!\!\!/_{D+1} = D\!\!\!\!/ -i \Gamma_y D_y$, $\Gamma_y^2 = 1$ 
 in which $D\!\!\!\!/$ is the $D$ dimensional covariant derivative, 
 $y$ is the $(D+1)$-th coordinate of the compactified space. 
The field strength for the gauge fields, the covariant derivative 
 and the triplet fermion are given by
\bea
F_{MN} &=& \partial_M A_N - \partial_N A_M -ig [A_M, A_N]~(M,N = 0,1,2,3, \cdots, D), \\
D_M &=& \partial_M -ig A_M~(A_M = A_M^a \frac{\lambda^a}{2}~(\lambda^a : \mbox{Gell-Mann~matrices})), \\
\Psi &=& (\psi_1, \psi_2, \psi_3)^T, 
\eea 
$g$ is the $(D+1)$ dimensional gauge coupling constant. 
We impose the periodic boundary conditions for a circle $S^1$ and $Z_2$ parity assignments as 
\bea
&&A_\mu = 
\left(
\begin{array}{ccc}
(+,+) & (+,+) & (-,-) \\
(+,+) & (+,+) & (-,-) \\
(-,-) & (-,-) & (+,+) \\
\end{array}
\right), \quad 
A_y=
\left(
\begin{array}{ccc}
(-,-) & (-,-) & (+,+) \\
(-,-) & (-,-) & (+,+) \\
(+,+) & (+,+) & (-,-) \\
\end{array}
\right), \nonumber \\ 
&&\Psi = 
\left(
\begin{array}{c}
\psi_{1L}(+,+) + \psi_{1R}(-,-) \\
\psi_{2L}(+,+) + \psi_{2R}(-,-) \\
\psi_{3L}(-,-) + \psi_{3R}(+,+) \\
\end{array}
\right)
\eea
where $(+,+)$ denotes that $Z_2$ parity are even at both fixed points at $y=0,\pi R$ for instance. 
 $R$ is the compactification radius. $\psi_{1L} \equiv 
 \frac{1}{2}(1-\gamma^y)\psi_1~(\Gamma^y \equiv i \gamma^y)$, etc. 
$\mu$ is a $D$-dimensional index ($\mu = 0, 1, 2, 3, \cdots, D-1$). 

This $Z_2$ parity assignments lead to the gauge symmetry breaking $SU(3) \to SU(2)_L \times U(1)_Y$, 
 the Higgs doublet is realized as the zero mode of $A_y$ and the chiral fermions are naturally obtained.

Substituting the KK mode expansions for gauge fields and a fermion satisfying 
 the above boundary conditions and integrating out $y$ coordinate, 
 we find the $D$-dimensional effective action of a fermion 
 relevant for the gluon fusion amplitude calculation 
\bea
&& S = \int d^D x \left[
\sum_{n=1}^{\infty} 
(\bar{\psi}_1^{(n)}, 
\bar{\tilde{\psi}}_2^{(n)}, \bar{\tilde{\psi}}_3^{(n)}) 
\right. \nonumber \\ 
&& \left. \times \left(
\begin{array}{ccc}
i \gamma^{\mu} \partial_{\mu} - m_{n} & 0 & 0 \\
0 & i \gamma^{\mu} \partial_{\mu} 
 -\left( m_{+}^{(n)} + \frac{m}{v} h \right) & 0 \\
0& 0 &i \gamma^{\mu} \partial_{\mu} 
 -\left( m_{-}^{(n)} - \frac{m}{v} h \right)  
\end{array}
\right)
\left(
\begin{array}{c} 
\psi_1^{(n)} \\
\tilde{\psi}_2^{(n)} \\
\tilde{\psi}_3^{(n)}
\end{array}
\right) \right. \nonumber \\ 
&& \left. + \ \mbox{gauge interaction part} + \ \mbox{zero-mode part} 
\right].  
\label{Ddimeff}
\eea
where this effective action is written in terms of KK fermion mass eigenstates 
 and the gauge interactions and zero mode terms are simply omitted 
 since they are irrelevant for the gluon fusion amplitude calculations. 
The detail derivation of this effective action is summarized in Appendix.

The essential point to show the finiteness is that 
 the structure of the mass eigenvalues and Yukawa couplings for KK mode fermions 
 seen in five dimensional case \cite{MO} is unchanged. 
Namely the mass splitting happens as $m_+^{(n)} = m_n + m$ 
 for $\tilde{\psi}_2^{(n)}$, $m_-^{(n)} = m_n - m$ for $\tilde{\psi}_3^{(n)}$ 
 where $m_n \equiv n/R$ is KK masses and $m$ is a fermion mass.  
Their Yukawa coupling is given by $-m/v$ for $\tilde{\psi}_2^{(n)}$, 
 $+m/v$ for $\tilde{\psi}_3^{(n)}$, respectively. 
The constant $v$ is a vacuum expectation value (VEV) of Higgs field.\footnote{In this paper, 
 the electroweak symmetry breaking is assumed to take place. 
Strictly speaking, we have to analyze the Higgs potential to examine 
 whether the electroweak symmetry breaking occurs or not. 
However, studying this issue is a hard task beyond the scope of this paper. 
Our obtained results are not affected by this assumption.} 
As will be seen later, the gluon fusion amplitude is expressed 
 as the difference between ``+'' mode contributions and ``$-$'' mode ones, 
 which cancels the divergence. 
This property is in sharp contrast to the universal extra dimension (UED) case \cite{UED} 
 where we have no KK mass splitting like $\sqrt{m_n^2 + m_t^2}$ 
 and Yukawa coupling is given by $-(m_t/v) \times (m_t/\sqrt{m_n^2 + m_t^2})$ \cite{Petriello}. 
Thus, the divergence cannot be in general canceled in the UED case.

Following these observations, 
 KK mode contributions to the gluon fusion amplitude is calculated, 
 which is just the $(D+1)$ dimensional extension to the result of \cite{Rizzo}, 
\bea
{\cal A} = -\frac{m}{v} g_s^2 \sum_{n=1}^\infty \int_0^1 dx \int_0^{1-x} dy 
\int \frac{d^Dk}{(2\pi)^D}
\left[
\frac{I_{\mu\nu}^{(n)+}}{2 [k^2+2kQ-(m_+^{(n)})^2]^3}  
- (+ \leftrightarrow -)
\right]
\label{amplitude1}
\eea
where $g_s$ is a QCD coupling, $x,y$ are Feynman parameters and 
\bea
I_{\mu\nu}^{(n) \pm} &\equiv& 2^{[D/2]} m_\pm^{(n)} 
[4k_\mu k_\nu +2(k_\mu p_\nu^2 - k_\nu p_\mu^1) 
- (p_\mu^1 p_\nu^2 - p_\mu^2 p_\nu^1) \nonumber \\
&&+ g_{\mu\nu}((m_\pm^{(n)})^2 - k^2 - p^1 \cdot p^2)], \\
Q_\mu &\equiv& y p_\mu^2 - x p_\mu^1.
\eea
$p^{1,2}$ denote external momenta of the gluons.

Making use of a formula
\bea
\frac{1}{D^s} = \int_0^\infty dt \frac{t^{s-1}}{\Gamma(s)}e^{-Dt}, 
\eea
we can rewrite ${\cal A}$ as
\bea
{\cal A} &=& -\frac{m}{v}g_s^2 \sum_{n=1}^\infty \int_0^1 dx \int_0^{1-x} dy 
\int \frac{d^Dk}{(2\pi)^D} \int_0^\infty dt \frac{t^2}{2 \Gamma(3)}
\left[ I_{\mu\nu}^{(n)+} e^{-[k^2 + 2kQ - (m_+^{(n)})^2]t} 
- (+ \leftrightarrow -) \right] \nonumber \\
&=& -\frac{m}{v} g_s^2\sum_{n=1}^\infty \int_0^1 dx \int_0^{1-x} dy 
\int_0^\infty dt \frac{t^2}{4(4\pi t)^{D/2}}
\left[ \hat{I}_{\mu\nu}^{(n)+} e^{[(m_+^{(n)})^2+Q^2]t} 
- (+ \leftrightarrow -) \right] 
\label{5}
\eea
where 
\bea
\hat{I}_{\mu\nu}^{(n)\pm} &\equiv& 
2^{[D/2]} m_\pm^{(n)} 
\left[ 
\tilde{I}_{\mu\nu} + \left(2-\frac{D}{2} \right) g_{\mu\nu}\frac{1}{t}
\right], \\
\tilde{I}_{\mu\nu} &\equiv& 4 Q_\mu Q_\nu 
+ 2(Q_\nu p_\mu^1 - Q_\mu p_\nu^2) -p_\mu^1 p_\nu^2 + p_\mu^2 p_\nu^1 
+ [(m_\pm^{(n)})^2 - (1 - 2xy) p^1 \cdot p^2] g_{\mu\nu} \nonumber \\
&=& -(1-2x)(1-2y)p_\mu^1 p_\nu^2 +(1-4xy) p_\mu^2 p_\nu^1 
+ (4x^2 - 2x)p_\mu^1 p_\nu^1 + (4y^2 - 2y)p_\mu^2 p_\nu^2 \nonumber \\
&& + [(m_\pm^{(n)})^2 - (1 - 2xy) p^1 \cdot p^2] g_{\mu\nu}
\eea
and the change of variable $k' = k + Q$ and the Gaussian momentum integral 
\bea
\int \frac{d^Dk}{(2\pi)^D} e^{-tk^2} = \frac{1}{(4\pi t)^{D/2}}, \quad 
\int \frac{d^Dk}{(2\pi)^D} k^2 e^{-tk^2} = \frac{D}{2t} \frac{1}{(4\pi t)^{D/2}}
\eea
was performed to arrive at the final expression (\ref{5}). 
Note that $\hat{I}_{\mu\nu}$ is symmetric under $x \leftrightarrow y$, 
 $\mu \leftrightarrow \nu$ and $p^1 \leftrightarrow p^2$ as it should be. 

As a consistency check, 
 focusing on the case with $D=4$, integrating with respect to $t$ 
 and using the relation $Q^2 = -xy m_H^2~(m_H: \mbox{Higgs mass})$ 
 and the fermion loop function
\bea
F_{1/2}(\tau) = -\tau \int_0^1 dx \int_0^{1-x} dy 
 \frac{4(1-xy)}{\tau - 4xy}~\left( \tau \equiv \frac{4(m_\pm^{(n)})^2}{m_H^2} \right),
\eea
we can verify that (\ref{5}) with the case $D=4$ agrees with the results in \cite{MO}. 

It is convenient for the demonstration of the finiteness 
 to include a half of zero mode contribution $(n=0)$ 
 because (\ref{5}) is further rewritten into the mode sum from $-\infty$ to $+\infty$ 
 as\footnote{Here we mean by a zero mode in a $D$ dimensional sense. 
In the present example, the zero mode contributions diverge in the case more than six dimensions. 
This is because the only one spatial dimension is compactified. 
In a realistic compactification, the zero mode contributions are always finite. 
Therefore, the divergence from the zero mode contributions is not a problem. 
Real problem is the divergences from KK mode contributions, 
 and what we will show is that this KK mode contributions indeed become finite 
 by showing that the full contribution is finite.}
\bea
{\cal A} &=& -\frac{m}{v} g_s^2 \sum_{n=-\infty}^\infty \int_0^1 dx \int_0^{1-x} dy 
\int_0^\infty dt \frac{t^2}{4(4\pi t)^{D/2}}
\hat{I}_{\mu\nu}^{(n)+} e^{[(m_+^{(n)})^2+Q^2]t} \nonumber \\
&=& -\frac{m}{v} g_s^2 \sum_{n=-\infty}^\infty \int_0^1 dx \int_0^{1-x} dy 
\int_0^\infty dt \frac{t^2}{4(4\pi t)^{D/2}}
2^{[D/2]} \left[ \tilde{I}_{\mu\nu} + \left(2-\frac{D}{2} \right)g_{\mu\nu}\frac{1}{t} 
\right] \nonumber \\
&& \times R^2 \sqrt{\frac{\pi}{t^3}}(i \pi n) 
e^{Q^2 t -\frac{(\pi Rn)^2}{t} -2\pi i n m R}.  
\label{Poissoned}
\eea
In the second equality, Poisson resummation formula is used. 
\bea
\sum_{n=-\infty}^\infty \left( \frac{n+a}{R} \right) 
e^{\left(\frac{n+a}{R}\right)^2 s} 
= \sum_{n=-\infty}^\infty R^2 \sqrt{\frac{\pi}{s^3}} (i \pi n) 
e^{-\frac{(\pi R n)^2}{s} -2\pi i na}
\label{Poisson}
\eea
which is just a Fourier transform from the KK momentum space 
 into the coordinate space of extra spatial dimensions. 
In other words, $n$ in the right-hand side of (\ref{Poisson}) 
 has a physical meaning of winding number around $S^1$. 

The divergence usually appears from no winding mode ($n=0$) in (\ref{Poissoned}) at $t=0$, 
 but this term trivially vanishes in the present case. 
Thus, the finiteness of the gluon fusion amplitude is verified.

Next, let us calculate the finite value explicitly. 
The finite value can be calculated from $n \ne 0$ part in (\ref{Poissoned}). 
Taking into account the gluon polarization sum 
$\epsilon^1 \cdot p^1 = \epsilon^2 \cdot p^2 = 0~(\epsilon^{1,2}: {\rm gluon~polarization~tensor})$ 
and doing the $t$ integral, we find
\bea
{\cal A} 
&=& -\frac{m}{v} g_s^2
\int_0^1 dx \int_0^{1-x} dy 
\frac{(1-4xy)p_\mu^2 p_\nu^1 R^2}{(4\pi)^{(D-3)/2}} (2^{[D/2]})(xy m_H^2)^{\frac{D-3}{4}}
\sum_{n=1}^\infty \frac{n \sin(2\pi nmR)}{(\pi Rn)^{\frac{D-3}{2}}} \nonumber \\
&& \times K_{\frac{D-3}{2}}(2\pi Rn m_H \sqrt{xy}) \nonumber \\
&\simeq&
-\frac{m g_s^2 p_\mu^2 p_\nu^1}{3 \cdot 2^{3(D-1)/2 - [D/2] } \pi^{3(D-3)/2} v R^{D-5}} 
\Gamma \left(\frac{D-3}{2} \right)
\sum_{n=1}^\infty \frac{\sin(2\pi nmR)}{n^{D-4}} 
\label{final}
\eea
where $K_\nu(x)$ is the modified Bessel function, 
\bea
\int_0^\infty dt t^{-\nu-1} e^{-At-B/t} = 
2 \left( \frac{A}{B} \right)^{\nu/2} K_\nu(2\sqrt{AB}). 
\eea
In order to obtain the final expression (\ref{final}), 
 the approximation formula for the modified Bessel function is used, 
\bea
K_{D/2}(x) \simeq x^{-D/2}2^{(D-2)/2}\Gamma(D/2)~(x \ll 1). 
\eea
(\ref{final}) is the final result of the gluon fusion 
amplitude.\footnote{For $D=4$, the mode sum becomes $\sum_{n=1}^\infty \sin(2\pi n mR) 
= \frac{1}{2}\cot(\pi mR)~(mR < 1)$. 
For $D=5$, $\sum_{n=1}^\infty \frac{\sin(2\pi n mR)}{n} 
= \frac{\pi}{2}(1-mR)$. These mode sums are certainly finite. 
As for other space-time dimensions, 
 the mode sum becomes trivially finite since the mode sum is bounded from the above 
 by at most $\sum_{n=1}^\infty 1/n^2=\pi^2/6$.} 
Although the finiteness was shown in this paper by using the model compactified on $S^1/Z_2$, 
 we empathize that the finiteness nature itself is not affected 
 by the shape of the compactified spaces. 
This is because the information of the compactification is an infrared property not an ultraviolet one. 
Off course, the finite value might be changed depending on the way of the compactification. 
If we consider more realistic compactifications, 
 the finiteness of the gluon fusion amplitude is trivial from the result in this paper 
 but it is not so trivial to calculate the finite value in such a realistic compactification. 
We are now working on the analysis of the gluon fusion 
 in a 6D gauge-Higgs unification model on $T^2/Z_4$ \cite{MO2}.

The finiteness of the gluon fusion amplitude can be also checked in a different way. 
Let us take first the mode sum before the momentum integration. 
The relevant mode sum is given by
\bea
&&\sum_{n=1}^\infty 
\left[
\frac{I_{\mu\nu}^{(n)+}}{[k^2+2kQ-(m_+^{(n)})^2]^3}
-
\frac{I_{\mu\nu}^{(n)-}}{[k^2+2kQ-(m_-^{(n)})^2]^3}
\right] 
\label{modesum}
\eea
which can be rewritten by including a half of zero mode contributions 
 as was done in the previous calculation, 
\bea
\sum_{n=-\infty}^\infty 
\frac{I_{\mu\nu}^{(n)+}}{[(k_E^2 + Q^2 + (m_+^{(n)})^2]^3}
\eea
where the change of variable $k'=k+Q$ is carried out  
 and $k_E$ denotes a Euclidean momentum.

The relevant mode sum can be classified into the following two types. 
\bea
\sum_{n=-\infty}^\infty \frac{4m_+^{(n)}A_{\mu\nu}(k_E,p)}{[k_E^2+Q^2+(m_+^{(n)})^2]^3}, \\
\sum_{n=-\infty}^\infty \frac{4g_{\mu\nu}m_+^{(n)}}{[k_E^2+Q^2+(m_+^{(n)})^2]^2}, 
\eea
where
\bea
A_{\mu\nu}(k_E,p) &=& 4(k_E)_\mu (k_E)_\nu + 4 Q_\mu Q_\nu - 2(Q_\mu p_\nu^2 - Q_\nu p_\mu^1) 
- (p_\mu^1 p_\nu^2 - p_\nu^2 p_\mu^1) \nonumber \\
&&+g_{\mu\nu}[-2Q^2 - p^1 \cdot p^2]. 
\eea
Making use of the formula, 
\bea
\sum_{n=-\infty}^\infty \frac{2(a+2n\pi)}{[x^2+(a+2n\pi)^2]^2} &=& 
\frac{\sinh x \sin a}{2x[\cosh x - \cos a]^2}, \\
\sum_{n=-\infty}^\infty \frac{4(a+2n\pi)}{[x^2+(a+2n\pi)^2]^3} &=& 
-\frac{\cosh x \sin a}{4x^2[\cosh x - \cos a]^2} \nonumber \\
&&+ \frac{\sinh x \sin a}{4x^3[\cosh x - \cos a]^2} 
+ \frac{\sinh^2 x \sin a}{2x^2[\cosh x - \cos a]^3}, 
\eea
we find that the relevant mode sums take the following forms in the large momentum limit, 
\bea
&&\sum_{n=-\infty}^\infty \frac{4m_+^{(n)}A_{\mu\nu}(k_E,p)}{[k_E^2 + Q^2 + (m_+^{(n)})^2]^3} 
\nonumber \\
&=& 4A_{\mu\nu}(2\pi k_E R, 2\pi pR) (2\pi R)^2 R \times \nonumber \\
&&\left[
-\frac{\cosh(2\pi R \sqrt{k_E^2+Q^2}) \sin(2\pi m R)}
{4(2\pi R)^2(k^2+Q^2)[\cosh(2\pi R \sqrt{k_E^2+Q^2}) - \cos(2\pi m R)]^2} 
\right. \nonumber \\
&& \left. +\frac{\sinh(2\pi R \sqrt{k_E^2+Q^2}) \sin(2\pi m R)}
{4(2\pi R)^3(k_E^2+Q^2)^{3/2}[\cosh(2\pi R \sqrt{k_E^2+Q^2}) - \cos(2\pi m R)]^2} 
\right. \nonumber \\
&& \left. +\frac{\sinh^2(2\pi R \sqrt{k_E^2+Q^2}) \sin(2\pi m R)}
{2(2\pi R)^2(k_E^2+Q^2)[\cosh(2\pi R \sqrt{k_E^2+Q^2}) - \cos(2\pi m R)]^3}
\right] \nonumber \\
&\to& -4 R (k_E)_\mu (k_E)_\nu \sin(2\pi mR) e^{-2\pi Rk_E}
\left(\frac{1}{k_E^2} + \frac{1}{2\pi Rk_E^3} \right)~(k_E \to \infty) \nonumber \\
&=& -4 \frac{R}{D} \left(1 + \frac{1}{2\pi Rk_E} \right) 
g_{\mu\nu} \sin(2\pi mR) e^{-2\pi Rk_E},
\label{UV1}
\eea
\bea
\sum_{n=-\infty}^\infty \frac{4g_{\mu\nu} m_+^{(n)}}
{[k_E^2 + Q^2 + (m_+^{(n)})^2]^2} 
&=& \frac{g_{\mu\nu}(2\pi R) \sinh(2\pi R \sqrt{k_E^2 + Q^2}) \sin(2\pi mR)}
{\sqrt{k_E^2 + Q^2}[\cosh(2\pi R \sqrt{k_E^2 +Q^2}) - \cos(2 \pi mR)]^2} \nonumber \\
&\to& g_{\mu\nu}(2\pi R)\frac{\sin(2\pi mR)}{k_E}e^{-2\pi Rk_E}~(k_E \to \infty). 
\label{UV2}
\eea
(\ref{UV1}) and (\ref{UV2}) immediately implies that the loop momentum integral 
 in any dimension becomes superconvergent. 
Namely, we have confirmed that the gluon fusion amplitude is also finite 
 by taking first the mode sum before the momentum integration.

This result can be understood by operator analysis in more general way. 
The local operator describing the gluon fusion 
 is given by the dimension six operator 
\bea
\langle H^\dag \rangle H G^a_{\mu\nu}G^{a\mu\nu} + h.c.
\label{operator}
\eea
where $H$ is the Higgs field and $G^a_{\mu\nu}$ is a field strength tensor for the gluons. 
In the gauge-Higgs unification, the Higgs is replaced by the extra component 
 of the higher dimensional gauge field $A_y$. 
Note that this $A_y$ cannot be written by the covariant derivative 
 $D_y$ as in the case of S and T parameters \cite{LM} 
 and the anomalous magnetic moment \cite{ALM}
 since the Higgs is neutral under $SU(3)_C$ gauge group. 
This implies that the operator (\ref{operator}) is forbidden 
 by the higher dimensional gauge invariance, 
 which leads to the finite result for the gluon fusion 
 in the gauge-Higgs unification. 
This argument holds ture for the brane localized operator of (\ref{operator}) 
 $\langle A_y^\dag \rangle A_y G^a_{\mu\nu}G^{a\mu\nu}$
 since the shift symmetry $A_y \to A_y + {\rm const}$ is operative even at branes \cite{GIQ}, 
 which is a remnant of the higher dimensional gauge symmerty. 
Therefore, the brane localized operator is forbidden by this shift symmetry. 
Furthermore, this operator analysis argument is very powerful 
 because this argument is model independent. 

In summary, 
 we have shown that the gluon fusion amplitude in the gauge-Higgs unification 
 is finite in any space-time dimension regardless of its nonrenormalizability. 
Taking a $D+1$ dimensional $SU(3)$ toy model of the gauge-Higgs unification with a triplet fermion 
 compactified on $S^1/Z_2$ to avoid inessential complications, 
 we have explicitly calculated the gluon fusion diagram 
 and verified its finiteness by two different ways of calculations. 
Note that the finiteness nature is independent of the shape of the compactified space 
 because the information on the compactification is the infrared property not the ultraviolet one. 
On the other hand, the finite value is affected by the compactification.

This result can be more generally understood by the operator analysis.  
The dimension six local operator describing the gluon fusion process 
 is forbidden by the higher dimensional gauge symmetry. 
The nonlocal finite term is generated by Wilson line effects. 
This operator analysis also holds true for the two photon decay process 
 since the Higgs field cannot be replaced by the covariant derivative for a photon field. 
Thus, we can expect the two photon decay to be finite as well as the gluon fusion. 

A few comments are in order. 
The first one is on the finiteness at higher order perturbations of the gauge coupling. 
In the second way of calculation taking first the mode sum before the momentum integration, 
 we have obtained that the gluon fusion amplitude is superconvergent. 
At the higher order perturbations, 
 the convergence property for the momentum integral becomes worse 
 due to the fact that the gauge coupling has a negative mass dimension 
 in the nonrenormalizable theory. 
However, the integrand of the momentum integral is exponentially suppressed 
 in the large momentum region. 
This suggests that the amplitude is finite even at any order of the perturbations 
 similar to the Higgs mass case \cite{2loop}. 
We note that we must renormalize order by order 
 the divergences arising from the subdiagrams 
 such as the gauge coupling corrections,  
 which cannot be forbidden by the higher dimensional gauge symmetry.
The finiteness beyond one-loop level can be shown to such an operator 
 with the renormalized gauge coupling as was done in Ref. \cite{2loop}.  

The second one is on the finiteness of the two photon decay of the Higgs boson amplitude 
 which is important mode at the LHC in the light Higgs boson case. 
The local operator describing the two photon decay amplitude is given by 
 $\langle H^\dag \rangle H F_{\mu \nu}F^{\mu \nu}$ 
 where $F_{\mu\nu}$ is the photon field strength. 
We can expect from this fact that the two photon decay amplitude is also finite 
 because the local operator is forbidden by the higher dimensional gauge symmetry in the following. 
In the two photon decay case, 
 the Higgs cannot be rewritten by the covariant derivative $D_y$ 
 to make a gauge invariant operator, $(D_y F_{\mu\nu}) (D^y F^{\mu\nu})$, 
 because of the vanishing commutator between $A_y$ and the photon part of $A_\mu^{(0)}$ 
 after the electroweak symmetry breaking $[\langle A_y \rangle, A_\mu^{(0)}]=0$. 
Thus, we can conclude that the two photon decay amplitude in the gauge-Higgs unification 
 is also finite as well as the gluon fusion amplitude.  

The third one is that our argument remains unchanged 
 in a realistic model allowing the top quark mass
 although the top quark is massless in our toy model. 
The difference between our toy model and a realistic one lies in 
   the representation of the fermions under the gauge group (see, for example \cite{CCP}). 
Our argument based on the operator analysis is independent of which representation 
   the fermion belongs to, which tells us that our results remain true even in a realistic model. 

The results obtained in this paper give an impact 
 on the LHC physics in the gauge-Higgs unification 
 since the gluon fusion amplitude is calculable and predictive in any dimension
 in spite of the fact that the theory is nonrenormalizable. 
On the other hand, 
 if we consider a theory of the UED with more than five dimensions, 
 the gluon fusion amplitude will diverge or depend on the cutoff scale of the theory. 
 which means that the results highly depends on the UV physics.  
Therefore, it is very interesting to calculate the gluon fusion amplitude 
 in the gauge-Higgs unification more than five dimensions. 
This issue is also phenomenologically interesting.  
The Higgs mass is generically predicted to be twice of $W$ boson mass 
 $m_H = 2 m_W$ from the tree level potential \cite{SSSW}, 
 which implies that the Higgs boson mainly decays into two $W$ bosons 
 by the standard model interaction. 
Namely, the effects of the gauge-Higgs unification are only contained in the gluon fusion amplitude.  
This leads to simplify the analysis greatly comparing to the two photon decay of Higgs boson. 
Following this observation, we are studying the gluon fusion process 
 in the 6D gauge-Higgs unification compactified on $T^2/Z_4$ \cite{MO2}.

\subsection*{Acknowledgments}
The author would like to thank C.S. Lim and N. Okada for useful discussions. 
He also would like to thank C.S. Lim for a careful reading of the manuscript. 
The work of the author was supported 
 in part by the Grant-in-Aid for Scientific Research 
 of the Ministry of Education, Science and Culture, No.18204024.

\appendix

\section{The derivation of $D$-dimensional effective action of a fermion}

In this appendix, we derive a $D$-dimensional effective action for a triplet fermion 
needed for the calculation of the gluon fusion amplitude in more detail. 
The action in $D+1$ dimensional we consider is simply given by 
\bea
{\cal L} = -\frac{1}{2} \mbox{Tr}  (F_{MN}F^{MN}) 
+ i\bar{\Psi}D\!\!\!\!/ \Psi
\label{lagrangian}
\eea
where $\Gamma^M=(\gamma^\mu, i \gamma^y)~(M = 0,1,2,3, \cdots, D; \mu = 0,1,2,3, \cdots, D-1)$,  
\bea
F_{MN} &=& \partial_M A_N - \partial_N A_M -i g_{5} [A_M, A_N], \\
D\!\!\!\!/ &=& \Gamma^M (\partial_M -ig_{5} A_M) \ \ 
(A_{M} = A_{M}^{a} \frac{\lambda^{a}}{2} \ 
(\lambda^{a}: \mbox{Gell-Mann matrices})),  \\
\Psi &=& (\psi_1, \psi_2, \psi_3)^T.
\eea
The periodic boundary conditions are imposed along $S^1$ for all fields. 
The non-trivial $Z_2$ parities are assigned for each field as follows, 
\bea 
\label{z2parity} 
A_\mu = 
\left(
\begin{array}{ccc}
(+,+) & (+,+) & (-,-) \\
(+,+) & (+,+) & (-,-) \\
(-,-) & (-,-) & (+,+) 
\end{array}
\right), \ \ 
A_y = 
\left(
\begin{array}{ccc}
(-,-) & (-,-) & (+,+) \\
(-,-) & (-,-) & (+,+) \\
(+,+) &(+,+) & (-,-)
\end{array}
\right), 
\eea
\bea 
\label{fermizero} 
\Psi = 
\left(
\begin{array}{cc}
\psi_{1L}(+,+) + \psi_{1R}(-, -) \\
\psi_{2L}(+,+) + \psi_{2R}(-, -) \\
\psi_{3L}(-,-) + \psi_{3R}(+, +) \\
\end{array}
\right),
\eea
where $(+,+)$ means that $Z_2$ parities are even at the fixed points $y=0$ 
and $y = \pi R$, for instance. $y$ is the $(D+1)$-th coordinate and 
$R$ is the compactification radius. 
$\psi_{1L} \equiv \frac{1}{2}(1-\gamma^y)\psi_1$, etc.

Following these boundary conditions, 
 KK mode expansions for the gauge fields 
 and the fermions are carried out.   
\bea
A_{\mu,5}^{(+,+)}(x,y) &=& \frac{1}{\sqrt{2 \pi R}} 
\left[
A_{\mu,5}^{(0)}(x) + \sqrt{2} \sum_{n=1}^\infty A_{\mu,5}^{(n)}(x) 
\cos \left( \frac{ny}{R} \right)
\right], \\
A_{\mu,5}^{(-,-)}(x,y) &=& \frac{1}{\sqrt{\pi R}} 
\sum_{n=1}^\infty A_{\mu,5}^{(n)}(x) 
\sin \left( \frac{ny}{R} \right), \\
\psi_{1L, 2L, 3R}^{(+,+)}(x,y) &=& \frac{1}{\sqrt{2 \pi R}} 
\left[
\psi_{1L, 2L, 3R}^{(0)}(x) 
+ \sqrt{2} \sum_{n=1}^\infty \psi_{1L,2L,3R}^{(n)}(x) 
\cos \left( \frac{ny}{R} \right)
\right], \\
\psi_{3L,1R,2R}^{(-,-)}(x,y) &=& i \ \frac{1}{\sqrt{\pi R}} 
\sum_{n=1}^\infty \psi_{3L,1R,2R}^{(n)}(x) 
\sin \left( \frac{ny}{R} \right). 
\label{KK}
\eea 
It is useful to introduce the overall factor $i$ in the last expansion 
 to make Yukawa coupling real after the chiral rotation performed later.

For the zero-mode of bosonic sector, 
 we obtain exactly what we need for the Standard Model: 
\bea
A^{(0)}_{\mu} = \frac{1}{2}
\left(
\begin{array}{ccc}
W^{3}_{\mu}+ \frac{B_{\mu}}{\sqrt{3}} & \sqrt{2} W^{+}_{\mu} & 0 \\
\sqrt{2} W^{-}_{\mu} & - W^{3}_{\mu}+ \frac{B_{\mu}}{\sqrt{3}} & 0 \\
0& 0 & -\frac{2}{\sqrt{3}} B_{\mu}   
\end{array}
\right) , \ \ 
A_5^{(0)} = \frac{1}{\sqrt{2}}
\left(
\begin{array}{ccc}
0 & 0 & h^{+} \\
0 & 0 & h^{0} \\
h^{-} & h^{0\ast} & 0 
\end{array}
\right),   
\eea
where $W_\mu^{3}, \ W_{\mu}^{\pm}$, $B_\mu$ are $SU(2)_L, U(1)_Y$ gauge fields
 and $h = (h^{+}, h^{0})^{t}$ is the Higgs doublet in the Standard Model. 
For the zero mode in the fermion sector, 
 a fermion corresponding to the right-handed top quark 
 $t_{R}$ is missing, which is irrelevant for demonstrating 
the finiteness of the gluon fusion amplitude. 
\bea
\Psi^{(0)} = 
\left(
\begin{array}{cc}
t_{L} \\
b_{L} \\
b_{R} \\
\end{array}
\right). 
\eea
The $SU(2)_{L} \times U(1)_{Y}$ gauge symmetry is broken 
 by the Higgs VEV, $\langle h^{0} \rangle = v/\sqrt{2}$, 
 in other words, $ \langle A_{5} \rangle = v/2 \; \lambda_{6}$.

After the gauge symmetry breaking, $D$-dimensional effective Lagrangian 
 among KK fermions, the Standard Model gauge boson and Higgs boson ($h$) 
 defined as $h^0=(v+h)/\sqrt{2}$ can be derived from the term 
 ${\cal L}_{{\rm fermion}}  = i\bar{\Psi}D\!\!\!\!/ \Psi$ 
 in Eq.~(\ref{lagrangian}). 
Integrating over the $(D+1)$-th dimensional coordinate, 
 we obtain a $D$-dimensional effective Lagrangian:  
\bea
{\cal L}_{{\rm fermion}}^{{\rm D-dim}} 
&=& 
\sum_{n=1}^{\infty}    
(\bar{\psi}_1^{(n)}, \bar{\psi}_2^{(n)}, \bar{\psi}_3^{(n)})    
\left(
\begin{array}{ccc}
i \gamma^\mu \partial_\mu - m_{n} & 0 & 0 \\
0 & i \gamma^\mu \partial_\mu - m_{n} & m + gh \\
0& m + gh  & i \gamma^\mu \partial_\mu - m_{n} 
\end{array}
\right)
\left(
\begin{array}{c} 
\psi_1^{(n)} \\
\psi_2^{(n)} \\
\psi_3^{(n)}
\end{array}
\right) \nonumber \\ 
&&  
+ {\rm gauge~interaction~part} 
+ {\rm zero-mode~part}
\label{Deffaction}
\eea
where 
 $m_{n} = \frac{n}{R}$ is the KK masses, 
 $g = \frac{g_{5}}{\sqrt{2\pi R}}$ is the 4D gauge coupling, 
 and $m = \frac{gv}{2} (= m_{W})$ is the bottom quark mass 
 in this toy model. 
In deriving the $D$ dimensional effective Lagrangian (\ref{Deffaction}), 
 chiral rotations 
\beq  
\psi_{1,2,3} \ \to \ e^{-i\frac{\pi}{4}\gamma^y} \psi_{1,2,3}  
\eeq
 have been made in order to get rid of $i \gamma^y$.  
The gauge interaction terms and the zero mode terms are simply neglected 
 because they are not needed to calculate the gluon fusion amplitude.

We easily see that the mass matrix for the KK modes 
can be diagonalized by use of the mass eigenstates 
$\tilde{\psi}_{2}^{(n)}, \ \tilde{\psi}_{3}^{(n)}$,   
\bea 
\pmatrix{ 
\psi_{1}^{(n)} \cr 
\tilde{\psi}_{2}^{(n)} \cr 
\tilde{\psi}_{3}^{(n)} \cr 
} 
= U 
\pmatrix{ 
\psi_{1}^{(n)} \cr 
\psi_{2}^{(n)} \cr 
\psi_{3}^{(n)} \cr 
}, \ \ \  
U =\frac{1}{\sqrt{2}}
\left(
\begin{array}{ccc}
\sqrt{2} & 0 & 0 \\
0 & 1 & -1 \\
0 & 1 & 1 
\end{array}
\right). 
\eea
In terms of these mass eigenstates for non-zero KK modes, 
 the Lagrangian is described as 
\bea
{\cal L}_{{\rm fermion}}^{{\rm D-dim}}  
&=& \sum_{n=1}^{\infty}  
(\bar{\psi}_1^{(n)}, \bar{\tilde{\psi}}_2^{(n)}, \bar{\tilde{\psi}}_3^{(n)}) 
\times \nonumber \\
&& \left(
\begin{array}{ccc}
i \gamma^{\mu} \partial_{\mu} - m_{n} & 0 & 0 \\
0 & i \gamma^{\mu} \partial_{\mu} 
 -\left( m_{+}^{(n)} + \frac{m}{v} h \right) & 0 \\
0 & 0 & i \gamma^{\mu} \partial_{\mu} 
 -\left( m_{-}^{(n)} - \frac{m}{v} h \right) \\
\end{array}
\right)
\left(
\begin{array}{c} 
\psi_1^{(n)} \\
\tilde{\psi}_2^{(n)} \\
\tilde{\psi}_3^{(n)} \\
\end{array}
\right) \nonumber \\ 
&&   + \mbox{gauge interaction part} + \mbox{zero-mode part}.  
\label{Deff}
\eea
The relevant Feynman rules for our calculation can be 
 read off from this Lagrangian.  
Note that the mass splitting $m_\pm^{(n)} \equiv m_{n} \pm m$ 
 occurs associated with a mixing 
 between the $SU(2)$ doublet component and singlet component. 
Furthermore, the mass eigenstate for $m_{+}^{(n)}$ 
 has Yukawa coupling $-m/v$, 
 while Yukawa coupling of the mass eigenstate for $m_{-}^{(n)}$ 
 has an opposite sign, $+m/v$. 
Together with the mass splitting of KK modes, 
 this property is a general result realized 
 in any gauge-Higgs unification model.

\end{document}